\documentclass[12pt]{elsart}
\usepackage{graphicx}
\usepackage{color}
\begin{document}
\begin{frontmatter}
\title{Production of Radioactive Nuclides in Inverse Reaction Kinematics}

\author[KVI]{E. Traykov},
\author[KVI]{A. Rogachevskiy},
\author[KVI]{U. Dammalapati},
\author[KVI]{P. Dendooven}
\author[KVI]{O.C. Dermois},
\author[KVI]{K. Jungmann},
\author[KVI]{C.J.G. Onderwater},
\author[KVI]{M. Sohani},
\author[KVI]{L.Willmann},
\author[KVI]{H.W. Wilschut \thanksref{HW}},
\author[Young]{A.R. Young}
\address[KVI]{Kernfysisch Versneller Instituut, Rijksuniversiteit Groningen, \\ Zernikelaan 25, 9747 AA Groningen, The Netherlands}
\address[Young]{Department of Physics,
NC State University, Box 8202 Raleigh, NC 27695 USA}
\thanks[HW]{Corresponding author,\\
Fax: +31-50-3633600, \\E-mail: wilschut@kvi.nl}

\begin{abstract}
Efficient production of short-lived radioactive isotopes in
inverse reaction kinematics is an important technique for various
applications. It is particularly interesting when the isotope of
interest is only a few nucleons away from a stable isotope. In
this article production via charge exchange and stripping
reactions in combination with a magnetic separator is explored.
The relation between the separator transmission efficiency, the
production yield, and the choice of beam energy is discussed. The
results of some exploratory experiments will be presented.

{\it PACS: 07.55.-w; 
           07.55.+h; 
                   29.30.-h; 
                   41.85.-p; 
                   41.75.-i; 
                   25.70.-z; 
        \\}

{\it Keywords: Magnetic separator, Inverse reaction kinematics,
Secondary radioactive isotopes
\\}

\end{abstract}

\end{frontmatter}

\section{Introduction}
Beams of radioactive nuclides can be produced in several ways.
Various strategies have been explored extensively, in particular
as starting points for the next generation of large scale
radioactive beam facilities such as FAIR \cite{FAIR}, RIA
\cite{Savard} and SPIRAL II \cite{SPIRAL}. A special case is the
production of nuclides that can be reached in simple
``rearrangement reactions'', such as charge exchange reactions,
e.g. (p,n) reactions, and stripping reactions, e.g. (d,p) and
(d,n) reactions. By inverting the kinematics of the reactions,
i.e. using a light particle target (H$_2$, D$_2$, $^{3,4}$He) and
a heavy-ion beam, secondary beams can be obtained with
considerable intensity and favorable emittance. Such beams have
been succesfully exploited, for example at Argonne National
Laboratories \cite{ANL} for astrophysical studies. At the
Cyclotron Institute of Texas A\&M the inverse reaction technique
is used at the MARS spectrometer \cite{MARS} for $\beta$-decay
studies. For the TRI$\mu$P facility at KVI Groningen, Netherlands,
these secondary beams will be used to study fundamental symmetries
and interactions via $\beta$-decay of light nuclei \cite{Wil}.

Recently, one key part of the facility, the TRI$\mu$P separator,
has become operational \cite{Berg}. The TRI$\mu$P separator
consists of two dipole sections. The first one separates the
primary and secondary beam, with the secondary beam dispersed in
the intermediate focal plane. The second section brings the
secondary beam back to an achromatic focus at the exit of the
separator. One can add an absorber in the intermediate plane,
which allows one to remove unwanted products that are produced
with the same rigidity as the desired secondary beam. In this
sense the TRI$\mu$P separator is like the magnetic separators used
for fragmentation reactions \cite{LISE,MSU}. In combination with
reactions in inverse kinematics clean and intense beams can be
obtained with such separators. Intensity limitations arise from
the separator acceptance both in momentum and angle.

In this article we explore in which way we can minimize the impact
of the separator acceptance. In particular, whether to choose low
or high energy beams. The various concepts to exploit inverted
reaction kinematics were tested experimentally. The methods were
applied already in the implantation of radioactive nuclides in Si
detectors, to study $\beta$-decay for fundamental and
astrophysical research \cite{LPC,Fynbo}.

\section{Kinematics}\label{kinematics}
\begin{figure}
\centering\includegraphics[width=40mm,angle=-90]{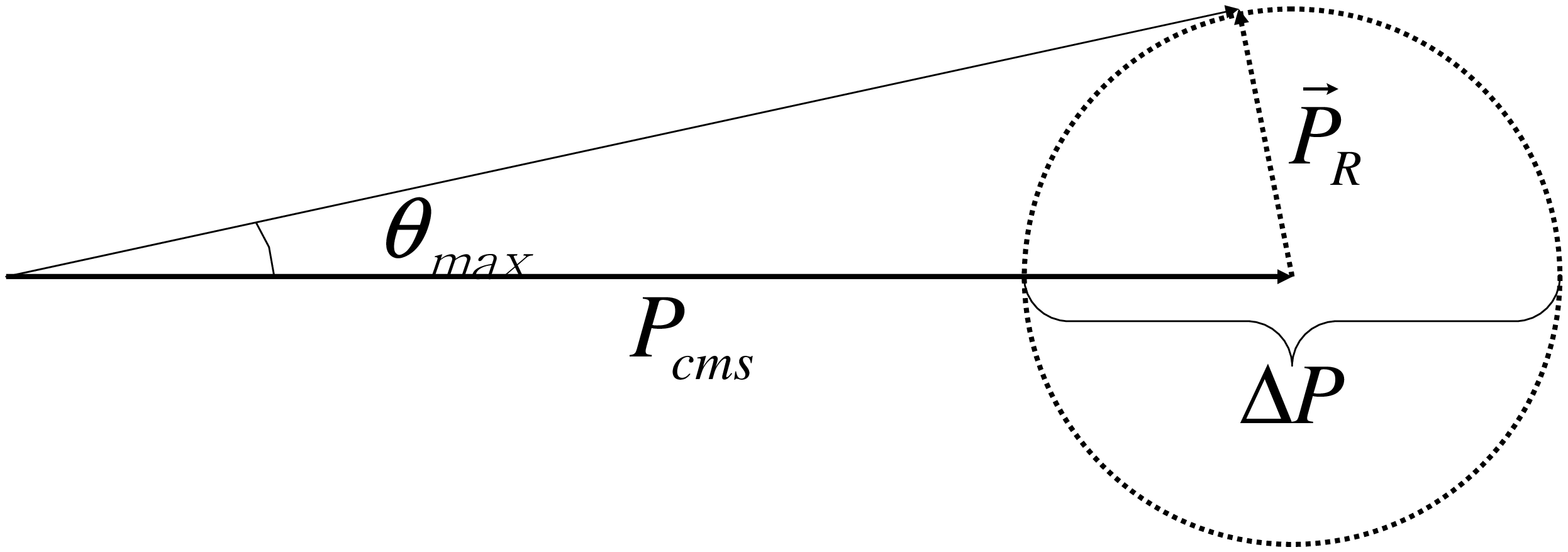}
\caption{\it Schematic diagram of the emittance of reaction
products in inverse kinematics. } \label{fig:vector}
\end{figure}

First we consider the momentum and angular acceptance. For the
purpose of our discussion it is sufficient to use non-relativistic
kinematics. The maximal variation in momentum, $\Delta p/p$ of the
projectile-like particle produced in a binary reaction is (see
figure \ref{fig:vector})
\begin{equation}\label{dpp} \frac{\Delta p}{p}=
\frac{2P_R}{P_{CMS}},
\end{equation}
where $P_{CMS}$ is the momentum associated with the center-of-mass
motion and $P_R$ is the recoil momentum  in the center of mass.
The projectile-like particles are emitted within a cone of
\begin{equation}\label{dth}
\Delta
\theta=2\theta_{max}=2\arcsin\frac{P_R}{P_{CMS}}\approx\frac{2P_R}{P_{CMS}}.
\end{equation}
Thus $\Delta\theta\approx\Delta p/p$. In terms of masses of the
projectile ($A_P$), target ($A_{T}$), projectile-like particle
($A_{PLF}$), and target-like particle ($A_{TLF}$)
\begin{equation}\label{dppq}
    \frac{\Delta p}{p}=2\sqrt{\frac{A_{TLF} A_T}{A_{PLF} A_P}(1 +
    \frac{Q}{E_{cm}})},
\end{equation}
where $Q$ is the energy excess (Q-value) of the reaction and
$E_{cm}$ is the center-of-mass energy. Thus in reactions where $
|Q|\ll E_{cm}$, such as in direct reactions, the momentum
distribution depends only on the masses. For example, in a (p,n)
reaction $\Delta p/p \approx 2/A_P$.  Another important result of
equation (\ref{dppq}) is that reactions with a large negative
Q-value will have small emittance. Moreover, when $-Q$ is much
larger than the Coulomb barrier in the entrance channel, the
reaction cross section nearly always peaks at threshold, mostly
because the reaction typically proceeds via a resonant state in
the compound nucleus.

Next we consider the angular distribution. We can write
\begin{equation}\label{dsigmadp}
\frac{d\sigma}{dp}
=\frac{d\sigma}{d\Omega}2\pi\sin\theta'\cdot\frac{p}{P_{CMS}P_R
\sin\theta'} ,
\end{equation}
where $\theta'$ is the center-of-mass angle and $p$ the laboratory
momentum of the projectile-like particle. When the center-of-mass
angular distribution $d\sigma/d\Omega$ is isotropic the resulting
momentum distribution is box shaped, i.e.
\begin{equation}\label{isotropic}
\frac{d\sigma}{dp}\propto\frac{p}{P_{CMS}P_R} \mbox{ for }
P_{CMS}-P_R\leq p \leq P_{CMS}+P_R.
\end{equation}
When the angular distribution approaches a $1/\sin\theta'$
dependence, singularities appear in the momentum distribution at
the limits $p=P_{CMS}\pm P_R$.

\section{Gas target}
For the production technique using inverse reactions a gas target
is essential. At high beam intensities a solid or liquid target of
e.g. Hydrogen would dissipate so much heat locally that it would
go into the gas phase. Also, a solid target with hydrogen, like
polyethylene, would loose rapidly its hydrogen content.

\begin{figure}
  \centering\includegraphics[width=7cm]{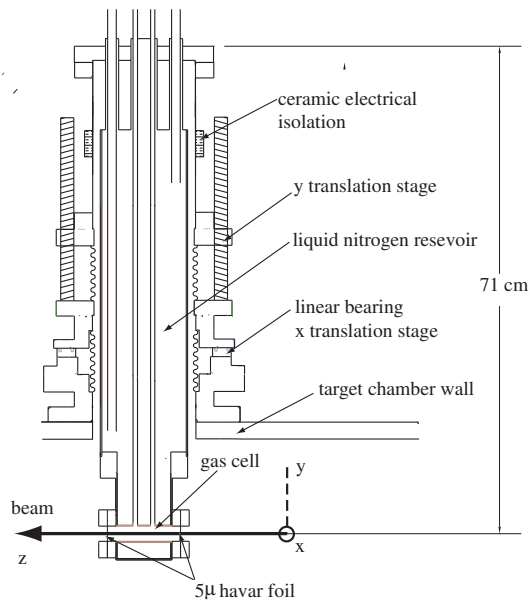}\\
  \caption{Schematic drawing of the gas target}\label{gastarget}
\end{figure}

The target for the TRI$\mu$P facility is based on a design from
Texas A\&M \cite{gasTAM} and has been built at North Carolina
State University \cite{Young}. The target has been used so far
with 1 bar of H$_2$ and with D$_2$. It is kept at liquid Nitrogen
temperature in a 10~cm long volume closed by windows of 5~$\mu$m
Havar of 1~cm diameter. The resulting target thickness is 3.2
mg/cm$^2$ for H$_2$. The beam diameter at the target position is
typically 2-3 mm. A schematic drawing of the target configuration
is shown in figure \ref{gastarget}.
The target can be moved remotely to allow the use of other targets
(e.g. a viewing target). The target is electrically isolated to
allow target-current read-out. The liquid-nitrogen dewar is
automatically filled. The target has been operated successfully in
all experiments so far, without replacing the windows.

\section{Reactions near threshold} We have measured
$d\sigma/dp$ for a number of reactions. Here we discuss results
for the reaction p($^{20}$Ne,$^{20}$Na)n at 22.3 MeV/nucleon
($E_{cm}=21.4$ MeV) with a Q-value of -14.7 MeV. The average
center-of-mass energy correcting for various energy losses is
20.7~MeV. This implies for this reaction that on average
$\Delta\theta=54$~mrad and $\Delta p/p=5.4$~\%, cf. equation
(\ref{dth}).  The emittance is thus close to the acceptance of the
separator, which is 60 mrad and 4\% in angle and momentum,
respectively. Had the Q-value been zero the emittance would have
been $\approx 2/A_P$ or 100 mrad and 10\%, respectively, i.e. 60\%
larger than for the actual Q-values. This illustrates the effect
of the large negative Q-value.

\begin{figure}
\centering\includegraphics[width=100mm,angle=0]{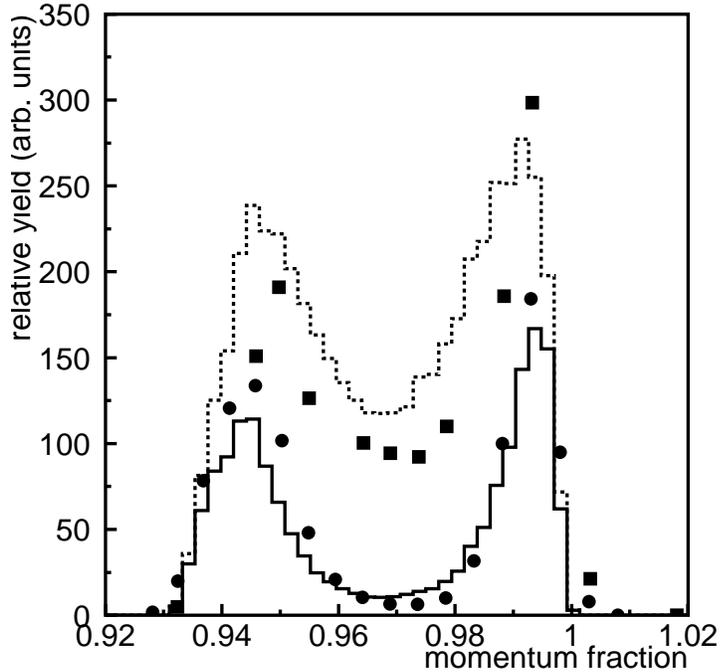}
\caption{\it Momentum dependence of the measured $^{20}$Na yield
for two angular openings of the separator. The filled circles
correspond to an angular opening of 32 mrad and the filled squares
to the maximum angular opening of 60 mrad. The histograms show the
result of Monte Carlo simulations assuming isotropic emission. The
full and dashed histogram refer to the small and large opening
respectively. The relative normalization of the two data sets is
uncertain due to systematic errors of about 20\%. }
\label{fig:dsdp}
\end{figure}

To measure $d\sigma/dp$ we scanned the rigidity of the separator,
using a Si detector with a 2 cm diameter in the intermediate focal
plane of the device. Two scans were performed: One with the full
angular opening and one with the angular acceptance restricted to
32 mrad. In figure \ref{fig:dsdp} we show the measured
$d\sigma/dp$ for these two acceptances.
For comparison we also give the results of Monte Carlo simulations
of an isotropic distribution that includes the effect of stopping
and straggling in the target and the beam divergence. The
calculations were normalized to the high momentum part of the scan
with a small angular opening. Systematic errors are about 20\%.
The data and calculations show that the momentum dependence of the
yield distribution can be well understood. The differences with
the data are in part due to the fact that the actual angular
distribution is not isotropic \cite{Bent} and part due to the
finite size of the detector.

\section{Direct reactions}
In the following we consider the effect of the angular
distribution on the kinematic focusing. Here we are concerned with
the shape of the angular distribution and its dependence on beam
energy. For (p,n) reactions this has been studied extensively and
a parametrization of the shape of the distribution is available
\cite{Tad1,Tad2}. The cross sections decrease approximately as
$1/E_{cm}$ well above threshold, while the angular distribution
becomes more focused at forward angles. We will argue below in
section \ref{thickness} that the target thickness can be increased
until the phase space of the separator is fully used, which allows
one  to compensate the $1/E$ dependence of the cross section.

The increasing acceptance with increasing energy does not stem
from the higher energy itself (see section \ref{kinematics}) but
from the fact that the angular distribution is peaked more
strongly around $0^\circ$. For (p,n) reactions this can be shown
explicitly. Here the angular distribution depends on the momentum
transfer, $q^2\approx k^2\sin^2\theta'/2$, with k the beam
momentum per nucleon. The cross section is approximated by
\cite{Tad1}
\begin{equation}
\frac{d\sigma}{dq}=\frac{2\pi}{k^2}\sigma_0 q \exp(-q^2 \langle
r^2\rangle/3), \label{cross section}
\end{equation}
where $\langle r^2\rangle$ is the mean square radius of the
projectile. The main point is that the cross section only depends
on the momentum transfer and that this holds at least
qualitatively over a large energy range. With this assumption the
secondary beam emittance decreases as $1-\exp(-q^2_{max}\langle
r^2\rangle/3)$, where $q_{max}$ is determined by $\theta_{max}$.
For other direct reactions a qualitatively similar emmitance
decrease with beam energy can be safely assumed. The measured
differential distribution of the yield as function of the rigidity
of the separator is
\begin{equation}\label{dsdq}
    \frac{d\sigma}{dp}=
    \frac{P_Rp}{P_{CMS}q}\frac{d\sigma}{dq}
    \propto\frac{P_Rp}{P_{CMS}}\exp(-q^2\langle r^2\rangle/3),
\end{equation}
where the last proportionality is only valid for (p,n) reactions.

\begin{figure}
  \centering \includegraphics[width=10cm]{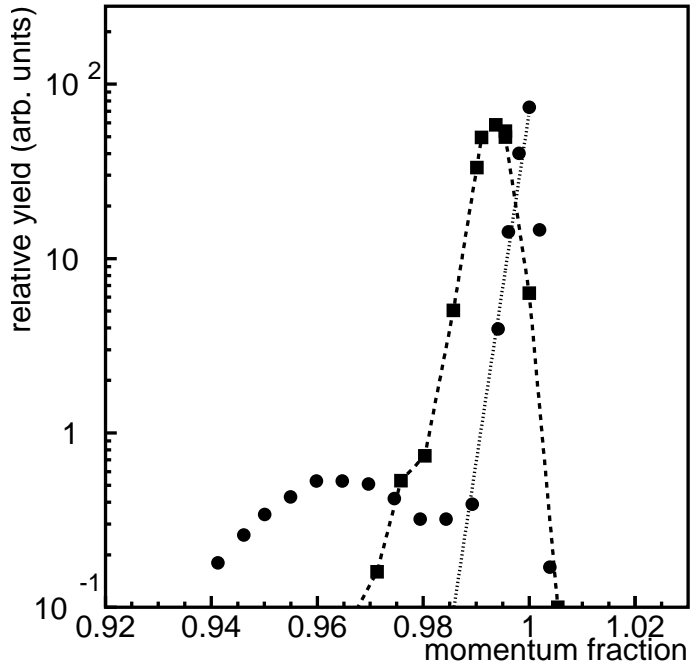}\\
  \caption{\it Momentum dependence of the reactions
  p($^{21}$Ne,$^{21}$Na)n at 43 MeV/nucleon
  and d($^{20}$Ne,$^{21}$Na)n at 22.3 MeV/nucleon.
  The result for the first reaction is indicated by the filled circles.
  The dotted line through the sharp rising part of the distribution is a fit using equation (\ref{dsdq}).
  The filled squares give the distribution of the stripping
  reaction; the dashed line connecting these data points is to guide eye.
  }\label{direct}
\end{figure}
We have measured the momentum dependence for two typical
reactions, the reaction p($^{21}$Ne,$^{21}$Na)n at 43 MeV/nucleon
and the reaction d($^{20}$Ne,$^{21}$Na)n at 22.3 MeV/nucleon. The
Q-value for these reactions are -4.3 and 0.2 MeV, respectively,
i.e. both reactions are well above threshold. The first reaction
has high yields because $^{21}$Ne and $^{21}$Na are mirror nuclei.
The production using the stripping reaction proceeds
preferentially via states with strong single particle nature.
Terakawa et al. \cite{Tera} found the largest population in the
first excited state of $^{21}$Na which corresponds to the $d(5/2)$
state.

The measured momentum distributions are presented in fig.
\ref{direct}. They were obtained in the same setup as described
above. The strong forward focus is very prominent in both
reactions. The main yield is in a momentum range of only 1 to 2
\%. For the (p,n) reaction the rising yield has been fitted with
equation (\ref{dsdq}) using $\langle r^2 \rangle$ as a free
parameter. A reduced radius $r_0=1.45$ fm was found, which is a
satisfactory result in view of the approximations and the
experimental method. The shape of the low yield tail for this
reaction is due to the combined effect of the changing acceptance
of the separator on one hand, and the characteristic oscillations
in the angular distributions at large angle on the other. Also the
stripping reaction shows a steep fall. In this case, however, the
energy straggling defines most of the shape of the distribution.
The window for the angular distribution extends only up to
$25^\circ$ in the center of mass. Nonetheless, it exhausts a large
fraction of the total yield as can be inferred from ref.
\cite{Tera}, which reports on angular distribution measurements at
the much lower energy of 12.5 MeV/nucleon.

\section{Target thickness}\label{thickness}
The yield in inverse kinematics can be maximized by increasing the
target thickness. However, at some point the added production will
fall outside the acceptance of the separator. One of the important
parameters is the stopping power. Because both beam and product
are slowed down, only the difference in stopping power needs to be
considered. One has exhausted the acceptance of the magnetic
separator when the thickness $d$ leads to a differential energy
loss
\begin{equation}\label{diffstop}
    \Delta
    E_{diff}=\left(1-\left(\frac{Z_{product}}{Z_{beam}}\right)^2\right)d\left(\frac{dE}{dx}\right)_{beam}
\end{equation}
that is larger than the energy acceptance of the separator. In
equation~(\ref{diffstop}) we have used the $Z^2$ dependence of the
stopping power $dE/dx$. Note that the entrance and exit windows
play no role. However, the amount of angular and energy
straggling, of course, does increase with the amount of material.

Integrating over the momentum dependence in equation (\ref{cross
section}) one obtains
\begin{equation}\label{integrated cross section}
    \sigma\approx\frac{3\pi}{k^2}\frac{\sigma_0}{\langle
    r^2\rangle}\approx 200\ \mbox{[MeV$\cdot$fm$^2$]}\frac{\sigma_o}{\mu^2\langle
    r^2\rangle\epsilon}\ ,
\end{equation}
where $\mu$ is the reduced mass in a.m.u. and $\epsilon$ is the
beam energy per nucleon. The maximum yield is then
\begin{equation}\label{Ymax}
    Y_{max}=I\sigma d=
    \frac{I (\sigma E) (\Delta E/E)}
    {\left(1-\left(\frac{Z_{product}}{Z_{beam}}\right)^2\right)\left(\frac{dE}{dx}\right)_{beam}},
    \end{equation}
where $I$ is the beam current and $\Delta E/E$ is the energy
acceptance of the separator (8\% for the TRI$\mu$P separator). In
the denominator of this equation the compensation of the energy
dependence of the cross section is made explicit. Moreover, the
stopping power in the nominator has also a $1/E$ dependence, in
this energy range. Thus if a production cross section is the
result of a direct reaction the highest beam energy should be
considered. For a (p,n) reaction only the constraints of the
accelerator are a limiting factor. Typical production rates of the
(p,n) and (d,p) reactions discussed were $3.2\times 10^3$ and $1.3
\times 10^4$~/s/particle-nA, respectively.

Using equation (\ref{Ymax}) to extrapolate towards higher yields
we find that increasing the target pressure 10-fold, eliminating
current constraints on the separator acceptance, and considering
the foreseen upgrade in beam current,  a production rate of $10^9$
radioactive particles/s can be reached with the TRI$\mu$P
separator in selected reactions.

\section{Conclusions}
We have considered two methods to produce efficiently a secondary
beam of radioactive particles. The methods are based on using a
light target and heavy-ion beam. The first method can be used in
reactions that have a large negative Q-value. One can select beam
energies near threshold, which results in small recoil energies,
while the products remain fast enough to allow to transmit and
separate them efficiently in a magnetic separator. The second
method exploits the increasingly narrow angular distributions.
Both methods can be roughly modelled so that these models can be
used to optimize the settings of the separator, using e.g. the
LISE code \cite{LISEC}. The methods described here can be used to
produce beams with rather narrow energy distributions, which can
be used as a starting point for radioactive beam experiments.
\section{Acknowledgments}
This work was supported by the \emph{Rijksuniversiteit Groningen}
and the \emph{Stichting voor Fundamenteel Onderzoek der Materie}
(FOM) under program 48 (TRI$\mu$P).
We thank the members of the AGOR cyclotron group and the KVI
support staff for their efforts.


\begin{thebibliography}{99}
\bibitem{FAIR} Conceptual design report, section 2, Gesellschaft f\"{u}r
Schwerionen Forschung, 2001; also
\verb+http://www.gsi.de/zukunftsprojekt/index_e.html+
\bibitem{Savard} G. Savard, in AIP-Conference-Proceedings  {\bf
656}, (2003) 335; also \verb+http://www.nscl.msu.edu/ria/+
\bibitem{SPIRAL}Report of the SPIRAL 2 Detailed Design Study
(APD), GANIL 2005; also
\verb+http://www.ganil.fr/research/developments/spiral2/whatisspiral2.html+
\bibitem{ANL} B. Harras et al., Rev. Sci. Instr. {\bf 71} (2000)
380.
\bibitem{MARS} R.E. Tribble, C.A. Gagliardi, and W. Liu, Nucl.
Instr. Meth. B {\bf 56/57} (1991) 956.
\bibitem{Wil} H.W. Wilschut, Hyperfine Interactions 146(2003)77;\\
H.W. Wilschut in AIP conference proceedings 802 (2005) 223; \\K.
Jungmann et al., Physica Scrypta T104(2003)178;\\M. Sohani, Acta
Physica Polonica B37(2006)231.
\bibitem{Berg} G.P.A. Berg et al., Nucl. Instr. and Meth. {\bf
A560} (2006) 169.
\bibitem{LPC} L. Achouri et al. in KVI annual report 2005 p.~13
\bibitem{Fynbo}S.G. Pedersen et al., to appear in Proceedings of Science (PoS)
and M.J.G. Borge et al. in KVI annual report 2005 p.~12.
\bibitem{LISE} D.Bazin et al.
Nucl. Instr. and Meth. A482 (2002) 307.
Nucl. Instr. and Meth. A 257 (1987) 215.
\bibitem{MSU} B.M. Sherrill, et al.
Nucl. Instr. and Meth. B 56/57 (1991) 1106.
\bibitem{gasTAM} J.F. Brinkley et al. in Cyclotron Institute of Texas
A\&M Universtiy Progress in Research, 2004 section V-9.
\bibitem{Young} A.R. Young et al. in KVI annual report 2004 p.17.
\bibitem{Bent} R.F. Bentley, thesis, Colorado University, 1972.
\bibitem{Tera} A. Terakawa et al., Phys. Rev. C 48 (1993) 2775.
\bibitem{Tad1} T.N. Taddeucci et al., Nucl. Phys. A469 (1987) 125.
\bibitem{Tad2} T.N. Taddeucci et al., Phys. Rev. C 41(1990)2548.
\bibitem{LISEC} O. Tarasov and D. Bazin, Nucl. Phys. A
746(2004)411.
\end{thebibliography}
\end{document}